\documentclass[10pt]{article} 
\usepackage{amssymb}
\usepackage{latexsym} 
\usepackage[dvips]{epsfig}

\def\ethbar{\overline{\eth}}

\def\sigmab{\overline{\sigma}} 

\def\cbar{\overline{c}} 
\def\Ybar{\overline{Y}}
\def\pbar{\overline{p}}

\def\pib{\overline{\pi}}

\def\alphab{\overline{\alpha}}
\def\mbar{\overline{m}}

\def\phib{\overline{\phi}} 

\def\phib{\overline{\phi}}

\def\Re{\mbox{Re}} 
\def\Im{\mbox{Im}} 
\def\O{\mbox{O}}
\def\i{\mbox{i}} 
\def\N{\mathcal{N}} 

\def\d{\mbox{d}}

\font\tenscr=rsfs10 scaled1100
\font\sevenscr=rsfs7 
\font\fivescr=rsfs5 
\skewchar\tenscr='177 
\skewchar\sevenscr='177 
\skewchar\fivescr='177
\newfam\scrfam 
\textfont\scrfam=\tenscr 
\scriptfont\scrfam=\sevenscr
\scriptscriptfont\scrfam=\fivescr

\def\scri{{\fam\scrfam I}}

\newcommand{\bs}{\hspace{-0.6cm}}

\begin{document}

\title{Boost-rotation symmetric type D radiative metrics in Bondi coordinates.}
\author{Ruth Lazkoz \thanks{E-mail address: {\tt  r.lazkoz@qmw.ac.uk
}}\\ Juan Antonio Valiente Kroon \thanks{E-mail address:  {\tt
j.a.valiente@qmw.ac.uk}} \\ School of Mathematical Sciences,\\ Queen
Mary \& Westfield College,\\ Mile End Road, London E1 4NS,\\ United
Kingdom.}

\maketitle

\begin{abstract}
The asymptotic properties of the solutions to 
the Einstein-Maxwell
equations with boost-rotation symmetry
and Petrov type D are studied. We find series solutions
to the pertinent set of equations which are suitable
for a late time descriptions in coordinates which
are well adapted for the description of the 
radiative properties of spacetimes (Bondi coordinates). By calculating the
total charge, Bondi and NUT mass and the Newman-Penrose constants
of the spacetimes we provide a physical 
interpretation of the free parameters of the solutions. Additional
relevant aspects on the asymptotics and radiative
properties of the spacetimes considered, such as
the possible polarization states of the gravitational 
and electromagnetic field, are discussed
through the way.
\end{abstract}

\section{Introduction}

Boost-rotation symmetric spacetimes  describe ``uniformly accelerated
particles'' that  approach the speed of light asymptotically. The
smoothness of the solution requires the spacetimes to be reflection
symmetric; therefore at least two particles with opposite acceleration
are present, and thus future null infinity contains at least two
singular points. Furthermore, these solutions are found to be time
symmetric, thus the puncturing of null infinity is present at both
$\scri^+$ and $\scri^-$. The null infinity of a boost-rotation
symmetric spacetime can be global, in the sense that it admits
spherical cuts, but the generators are not complete \cite{BicSch89a}.

This family of spacetimes possess an Abelian $G_2$ of isometries, in
which one of the Killing vector fields has closed orbits, and the
other symmetry is the curved spacetime generalization of the boost
rotations of Minkowski spacetime that leave invariant the null cone
with vertex at the origin \cite{BicSch84}. The boost-rotation symmetry
is of special interest because it is the only other continuous
isometry an axially symmetric spacetime can have that does not exclude
the possibility of radiation \cite{BicSch84}. Bi\v{c}\'{a}k,
Hoenselaers \& Schmidt have shown how to systematically construct an
infinite number of these spacetimes by prescribing the multipolar
structure of the particles undergoing the hyperbolic motion
\cite{BicHoeSch83a,BicHoeSch83b}.

One of the most well known representatives of the family of
boost-rotation symmetric spacetimes is the C-metric, a type D
spacetime with hypersurface orthogonal Killing vectors which has been
interpreted as describing a pair of black holes receding from one
another, and joined by a singular axis \cite{KinWal70,CorUtt95a}. The
C-metric can be generalized to include Maxwell field with principal
null directions aligned with the pairs of repeated principal null
directions of the Weyl tensor. Early works on boost-rotation symmetric
spacetimes demanded the Killing vector fields to be hypersurface
orthogonal, but as discussed by Bi\v{c}\'{a}k and Pravdov\'{a}
\cite{BicPra98}, this hypothesis can be set aside. Thus, another
possible  generalization is the twisting C-metric \cite{PleDem76}. In
the charged version it is found that the two accelerated black holes
have opposite charge so that the overall spacetime is electrically
neutral \cite{CorUtt95b}. Ashtekar \& Dray \cite{AshDra81} have shown
that the C-metric admits a conformal completion such that the cuts of
$\scri$ are isomorphic to the 2-sphere, and therefore it can be regarded
as asymptotically flat at null infinity. Subsequent work by Dray
showed that the charged C-metric is also asymptotically flat at
spatial infinity \cite{Dra82}. He also showed that the symmetries of
the spacetime give rise to a vanishing ADM mass. 

The boost-rotation symmetric spacetimes with hypersurface orthogonal
Killing vectors  are usually given in a coordinate system that clearly
exhibits their symmetries \cite{BicSch89a}. In these symmetry-adapted
coordinates, the metric of the spacetime contains only two
functions. The different particular examples can be obtained from
boost-rotation symmetric solutions of the flat spacetime wave equation
with sources. Once one of these is prescribed, the metric functions
can be obtained by quadratures. If one wants to study the radiative
properties of these spacetimes (radiation patterns, mass loss,
Newman-Penrose constants), then one has to rewrite the spacetime in
terms of Bondi coordinates like the ones discussed in chapter 3.  The
transformation between the symmetry adapted coordinate system and the
Bondi coordinates used in the asymptotic expansions of the
gravitational field  has to be given in terms of series
\cite{Bic68}. These expansions are extremely messy, and usually only
the leading terms can be calculated explicitly \cite{Bic68}. To add to
the problem, the coefficients in terms of which some quantities of
interest like the Newman-Penrose constants are defined are found deep
into the series expansions. One is bound to look for better methods to
calculate the quantities of physical interest.

Bi\v{c}\'{a}k and Pravdov\'{a} \cite{BicPra98} have obtained a series
of constraint equations for the news function and the mass aspect of
electrovacuum boost-rotation symmetric spacetimes.  The constraint
equations can be solved so that the news and the mass aspect depend on
arbitrary functions of the ratio $w=\sin\theta/u$. These functions
have, of course, to satisfy the field equations. Therefore, if one
makes some extra assumptions about the spacetime one can obtain a
closed system of ordinary differential equations for the mass aspect,
the gravitational and electromagnetic news function.

Taking the C-metric as paradigm, our attention will be restricted to
type D spacetimes. For electrovacuum spacetimes it will be necessary
to make a further assumption about the principal null directions of
the electromagnetic field; it will be taken to be algebraically
general, but with null principal directions parallel to the pairs of
null principal directions of the Weyl tensor. In order to be able to
make some discussion on the polarization of the gravitational and
electromagnetic radiation fields, the Killing vector fields will not
be assumed hypersurface orthogonal. The family of boost-rotation
spacetimes that will come out of these assumptions will clearly
contain the C-metric as a particular case: hypersurface orthogonal
Killing vectors, and no electromagnetic field.
 
Some analysis of the radiative properties of the spinning, charged
C-metric of Plebanski \& Demianski \cite{PleDem76} was done by
Farhoosh \& Zimmerman \cite{FarZim80a,FarZim80b}. They wrote the
spacetime in terms of Bondi coordinates and evaluated some radiative
properties of the spacetime (the news function and the mass aspect) in
the linear regime. However, they failed to notice the global character
of the solution, and thus for example they did not find that the
overall electromagnetic charge of the spacetime vanishes. More
recently, there has been some work on the spinning C-metric by
Bi\v{c}\'{a}k \& Pravda \cite{BicPra99} in which they discuss briefly
the radiative character of this spacetime, and show how it indeed
describes two uniformly accelerated, spinning black holes connected by
a conical singularity, or with conical singularities extending from
each horizon to infinity.

At this point it is worth making a note about the radiative character
of the C-metric in particular, and the boost-rotation symmetric
spacetimes as a whole. A boost-rotation symmetric spacetime can be
divided in three regions depending on whether the Killing vector is
timelike, null or spacelike. The region where the Killing field is
null is known as the \emph{roof}, and it intersects $\scri$ at the
cuts where the particles in hyperbolic motion puncture null
infinity. Below the roof, the Killing vector is timelike, and thus the
spacetime is stationary in this region. Above the roof, the Killing
vector field is timelike, so that the spacetime can be radiative. In
this region, the boost-rotation symmetric spacetimes can be shown to
be locally isometric to cylindrical waves (Einstein-Rosen waves)
\cite{BicSch89a}. It is also important to note that these solutions
have a time reflexion symmetry. Thus, what is true for $\scri^+$ is
also true for $\scri^-$. Hence, the particles come from past null
infinity, leaving a puncture on it. They approach to each other, and
then they recede again. Finally, they again puncture null infinity
(but now at $\scri^+$).

Previous attempts at addressing the radiative  properties of
boost-rotation symmetric spacetime proved to be unfruitful, mainly
because they relied on finding a transformation between the symmetry
adapted coordinates and the Bondi coordinates. Unfortunately, it is
not  possible in general  to express the coordinate transformation in
a closed form. This difficulties have motivated  us to adopt a rather
different approach here, as we construct the solutions by using series
expansions. Here we take advantage of the  series of constraint
equations for the news function and the mass aspect of electrovacuum
boost-rotation symmetric spacetimes found elsewhere
\cite{BicPra98,Val00b}.  These equations are the tools which will help
us  gain some insight in the structure and properties of the
Newman-Penrose constants of  the mentioned class of radiative
spacetimes. Only those solutions to the constraint equations can be
solved  such that the news and the mass aspect depend on an  arbitrary
function of the ratio $w=\sin\theta/u$ are considered here. Needless
to say that these  functions will have to  satisfy the field
equations. Under the assumptions made, we obtain a closed system of
ordinary differential equations for the mass aspect, the gravitational
and electromagnetic news function. With these at hand we are able to
describe and interpreting the asymptotic and radiative properties of
the spacetimes under consideration, in what their late time limit is
regarded. We give expressions for the total charge, Bondi and NUT
masses and the Newman-Penrose constants and with these at hand  we
succeed in attaching a physical meaning to the main parameters
defining the solutions. We also give some additional arguments which
provide an independent backup to our interpretation.  Then, we carry
out a counting of the  degrees of freedom of  both the gravitational
and electromagnetic field. Finally, we outline our main conclusions.

\begin{figure}[tbt]
\centering \mbox{ \epsfig{file=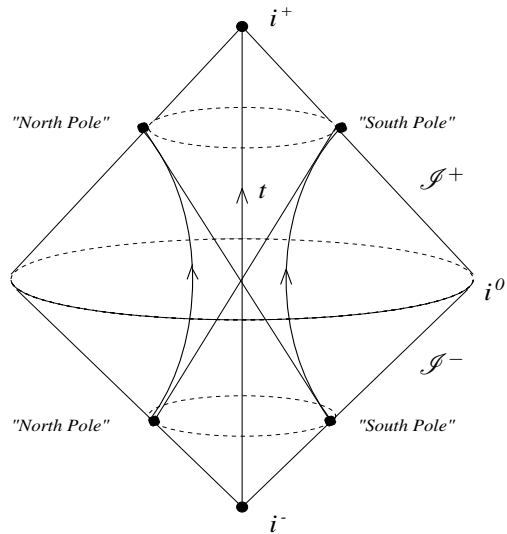,width=10cm,height=7cm}}
\put(-90,130){$\scri^+$} \put(-90,60){$\scri^-$}
\caption{Penrose diagram of a boost rotation symmetric spacetime. The
darker curves represent the world line of the particles undergoing
hyperbolic motion.}
\end{figure}

\section{Constraints}

\subsection{Asymptotic flatness and Bondi coordinates.}

To begin with, consider
a foliation of an asymptotically flat spacetime by ---say--- future
 oriented null hypersurfaces. Let this foliation be parametrized by
a retarded time coordinate $u$. On each null hypersurface one takes a generic
 geodesic generator as a  null curve parametrized by an
 affine parameter $r$. At a particular cut of $\scri$ (the
 intersection of the null hypersurfaces with null infinity),
coordinates 
$x^\alpha$ are introduced, where $\alpha=2,3$. These will be
 taken to be the usual angular coordinates $(\theta,\varphi)$, 
although a complex
 stereographic coordinate $\zeta$ could be used as well. In addition,
these coordinates can be propagated 
along $\scri$ and into the interior of the spacetime
by requiring them to remain constant along the generators of $\scri$
and  the outgoing null geodesics respectively.
This coordinate construction will be referred as to
 \emph{Bondi coordinates}. 

In order to construct a null tetrad, one takes as its first
vector $l^a$  the vector tangent  
the null hypersurfaces giving the foliation. The scaling in the
affine parameter $r$ is then chosen so that the null vector coincides with
$\partial _r$. Now, by looking at the 2-surfaces $u=const.$, $r=const.$
($S_{u,r}$) it can be seen that, at each point on these, there is another
null vector $n^a$ which is future pointing and orthogonal to the
2-surface. Then, $n^a$ will be chosen as the second vector of the
tetrad. Finally, the vectors $m^a$ and $\mbar^a$ are chosen so that they span
the tangent space of $S_{u,r}$. From the previous construction it can
be seen that:
\begin{eqnarray}
&&l^a=\partial_r, \\
&&n^a=\partial_u+Q\partial_r+C^\alpha\partial_\alpha, \\
&&m^a=\xi^a\partial_\alpha,
\end{eqnarray}
with $\alpha=\theta$, $\varphi$. The remaining freedom in the tetrad
construction consists on a boost ($l\mapsto Al$, $n\mapsto
A^{-1}n$), which can be used to rescale $r$; and a spin ($m\mapsto
e^{i\theta}m$), which in turn can be used in to make the spin coefficient
$\epsilon$ vanish. It can be also shown that
$\kappa=0$, $\tau=\pib=\alphab+\beta$, and $\rho$ and $\mu$ are real.

The spacetimes under consideration will be assumed to be
asymptotically flat with smooth sections of $\scri$, which is
equivalent to say that the Weyl tensor of the spacetime peels off. 
Similarly, we will assume that the electromagnetic field also peels off. Thus,
\begin{eqnarray}
\Psi_n= {\cal O}(r^{n-5}), \\ \phi_m={\cal O}(r^{m-3}),
\end{eqnarray}
with $n=0\ldots4$, $m=0\ldots2$. The asymptotic expansions in powers
of $1/r$ for peeling-off asymptotically flat Einstein-Maxwell systems
are well known, and can be found in several places in the literature
\cite{PenRin86,Ste91}. Here Stewart's version \cite{Ste91} will be
used \footnote {It has to be pointed out that these expansions differ
slightly from those appearing in Penrose \& Rindler \cite{PenRin86},
as the tetrads used there are also different.}. Explicitly, from the
peeling-off theorem one has:
\begin{eqnarray}
&&\Psi_n=\Psi_n^{n-5}r^{n-6}+\Psi_n^{n-6}r^{n-6}+{\cal O}(r^{n-7}),
\label{peel1}\\
&&\phi_m=\phi_m^{m-3}r^{m-3}+\phi_m^{m-4}r^{m-4}+{\cal O}(r^{m-5})
\label{peel2}
\end{eqnarray}
with $n=0\ldots4$, $m=0\ldots2$.  It also follows
$\sigma=\sigma_2r^{-2}+{\cal O}(r^{-3})$.  In general, the
coefficients in the  expansions (\ref{peel1})-(\ref{peel2}) will depend
on $(u,\theta,\varphi)$. The coefficients in $\Psi_0$ ($\Psi_0^5$,
$\Psi_0^6,\ldots$) are the initial data quantities over a null
hypersurface $\N_0$ in a characteristic initial value problem, whereas
$\Psi_1^4$ and $\Re\,\Psi_2^3$ (the Coulomb part of the gravitational
field) are data that have to be supplied at $\N_0\cap\scri^+$. A
similar thing happens with the Maxwell field: $\phi_0$ ($\phi_0^3$,
$\phi_0^4,\ldots$) being data at $\N_0$ and $\phi_1^2$ (the Coulomb
part of the Maxwell field) being data at $\N_0\cap\scri^+$. The
remaining data is set by supplying  $\phi_2^1$ and the leading term
of $\sigma$. The coefficient $\sigma_2$ determines the radiative part
of the gravitational field via:
\begin{eqnarray}
&&\Psi_4^1=-\ddot{\sigmab}_2, \\  &&\Psi_3^2=-\eth \dot{\sigmab}_2.
\end{eqnarray}
The imaginary part of $\Psi_2^3$ is related to $\sigma_2$ through:
\begin{equation}
2\,\Im(\Psi_2^3)= \ethbar^2\sigma_2-\eth^2\sigmab_2 +
\sigmab_2\dot{\sigma}_2-\dot{\sigmab}_2\sigma_2.
\end{equation}
In addition, the following relations will be required later:
\begin{eqnarray}
&&\Psi_1^5=6\phib_1^2\phi_0^3-\ethbar\Psi_0^5, \\
&&\Psi_2^4=4\phi_1^2\phib_1^2-\ethbar\Psi_1^4, \\
&&\Psi_3^3=2\phib_1^2\phi_2^1-\ethbar\Psi_2^3, \\
&&\Psi_4^2=-\ethbar\Psi_3^2,
\end{eqnarray}
together with,
\begin{eqnarray}
&&\phi_1^3=-\ethbar\phi_0^3, \\
 &&\phi_2^2=-\ethbar\phi_1^2\,.
\end{eqnarray}
These relation are obtained from the expansions of the $D$-Bianchi identities and Maxwell equations. Similarly, from the $\Delta$-Bianchi identities one obtains the evolution equations (i.e. the equations with derivatives with respect to $u$):
\begin{eqnarray}
&&\dot{\Psi}_0^5=\eth \Psi_1^4 + 3\sigma_2 \Psi_2^3
+6\phi_0^3\phib_2^1,\label{psi05} \\
&&\dot{\Psi}_0^6=-4\ethbar(\sigma_2\Psi_1^4)-
\ethbar\eth\Psi_0^5+8\phib_1^2\eth\phi_0^3-
16\sigma_2\phi_1^2\phib_1^2+8\phi_0^4\phib_2^2,\label{psi06} \\
&&\dot{\Psi}_1^4=\eth \Psi_2^3 + 2\sigma_2 \Psi_3^2
+4\phi_1^2\phib_2^1,\label{psi14} \\ &&\dot{\Psi}_2^3=\eth \Psi_3^2
+ \sigma_2\Psi_4^1 +2\phi_2^1\phib_2^1, \label{psi23}
\end{eqnarray}
and
\begin{eqnarray}
&&\dot{\phi}_0^3=\eth\phi_1^2+\sigma_2\phi_2^1,\label{phi03} \\
&&\dot{\phi}_0^4=-\ethbar\eth\phi_0^3-
2\ethbar(\sigma_2\phi_1^2),\label{phi04} \\
&&\dot{\phi}_1^2=\eth\phi_2^1\label{phi12}.
\end{eqnarray}

\subsection{The algebraic type of the gravitational and electromagnetic fields}
As remarked in the introduction, we make extra assumptions
are required in order to obtain a closed system of ordinary
differential equations. Our analysis will be restricted to type D
spacetimes (i.e. spacetimes with two pairs of repeated null principal
directions). A type D spacetime is obtained by demanding  the
quartic
\begin{equation}
\Psi_0-4c_w\Psi_1+6c^2_w\Psi_2-4c_w^3\Psi_3+c^4_w\Psi_4=0, \label{pnd}
\end{equation}
to have two different pairs of repeated roots. This is the case if:
\begin{eqnarray}
&\!&\Psi_0=(\Psi_4)^{-3}\left[\,3\,\Psi_2\Psi_4-2\,(\Psi_3)^2\,\right]^2, 
\label{typed1}
\\
&\!&\Psi_1=\Psi_3\,(\Psi_4)^{-2}\left[3\,\Psi_2\Psi_4-2\,(\Psi_3)^2\right], 
\label{typed2}
\end{eqnarray}
provided that $(\Psi_3)^2\ne\Psi_2\Psi_4$, otherwise one gets a type N
spacetime\cite{KSMH}. From these two relations one obtains expressions for the
coefficients in the expansions of $\Psi_0$ and $\Psi_1$
(i.e. $\Psi_0^5$, $\Psi_0^6$, $\Psi_1^4$). The relations are:
\begin{eqnarray}
\Psi_0^5=&&\bs{{(\Psi_4^1)^{-3}}\left[\,3\,\Psi_2^3\Psi_4^1-2\,(\Psi_3^2)^2\,\right]^2},
\\
\Psi_0^6=&&\bs {(\Psi_4^1)^{-4}}{\left[2(\Psi_3^2)^2-3\Psi_4^1\Psi_2^3\right)
\left(3\,\Psi_4^2(\Psi_4^1\Psi_2^3-2(\Psi_3^2)^2) -6\Psi_2^4(\Psi_4^1)^2
+8\Psi_4^1\Psi_3^2\Psi_3^3\right]},
\\
\Psi_1^4=&&\bs{(\Psi_4^1)^{-2}}{\Psi_3^2\,\left[3\,\Psi_2^3\Psi_4^1-2\,
(\Psi_3^2)^2\right]}{(\Psi_4^1)^{-2}}.
\end{eqnarray}
The class of electromagnetic fields to be considered are such that
their principal null directions are aligned with the two pairs of
repeated principal null directions of the Weyl tensor. This
requirement  can be implemented as follows. The condition for a null
rotation  (about $n^a$) to give $l^a$ pointing in a principal null
direction is given by equation (\ref{pnd}). If the spacetime is of
type D (i.e. equations (\ref{typed1}) and (\ref{typed2}) hold), then
the double repeated solutions of  this quartic equation are:
\begin{equation}
c_w=\frac{\Psi_3 \pm
\sqrt{3\,\left((\Psi_3)^2-\Psi_2\Psi_4\right)}}{\Psi_4}.
\end{equation}
The analogous condition for the Maxwell field is given in terms of
 a quadratic equation,
\begin{equation}
\phi_0-2c_m\phi_1+c^2_m\phi_2=0,
\end{equation}
the roots of which are given by
\begin{equation}
c_m=\frac{\phi_1 \pm \sqrt{{(\phi_1)^2-\phi_0\phi_2}}}{\phi_2}.
\end{equation}
Demanding the roots of the quartic and quadratic equations to be equal
 by pairs, so that the principal null
directions of the Maxwell field are aligned with the pairs of repeated
principal null directions of the Weyl tensor, one arrives at the
condition:
\begin{eqnarray}
&&{\phi_0}={\phi_2}\left(3{\Psi_2}{\Psi_4}^{-1}-2{(\Psi_3)^2}
{(\Psi_4)^{-2}}\right),
\end{eqnarray}
from which it can be seen that:
\begin{eqnarray}
\bs\bs&&\phi_0^3=\phi_2^1\left(3{\Psi_2^3}{\Psi_4^1}^{-1}-
2{(\Psi_3^2)^2}
{(\Psi_4^1)^{-2}}\right),\\
\bs\bs&&\phi_0^4=2\,(\Psi_4^1)^{-3}\phi_3^2\left(\phi_3^2 
\left(2\,\phi_4^2\Psi_2^1 - 
     \phi_4^1\Psi_2^2 \right)-
2\,\phi_3^3\phi_4^1\Psi_2^1\right)  + 
  3\,\phi_4^1\left(\phi_2^4\phi_4^1\Psi_2^1 - 
     \phi_2^3\left( \left( \phi_4^2{\Psi}_2^1
            \right)  - \phi_4^1{\Psi}_2^2 \right) 
      \right),
\end{eqnarray}

\subsection{The Killing vector fields}

The spacetimes under consideration will be assumed to 
have two Killing vector fields ($\xi_1^a$ and $\xi_2^a$). 
The Killing vector $\xi_1^a$ will be taken to be an axial
vector field (closed orbits). The Bondi coordinates can be
 chosen so that $\xi_1^a=\partial_\varphi$. Because of 
the closed orbits of $\xi_1^a$, the $G_2$ will be necessarily Abelian 
\cite{BicSch84}.

It was shown elsewhere \cite{Val00b} that
 the asymptotic form of the most general Killing 
vector field compatible with asymptotic flatness is given by:
\begin{equation}
\xi^a=\left( -a_{-1}u+{\cal O}(1/r), 
a_{-1}r+{\cal O}(1), -\frac{1}{\sqrt{2}}(\cbar_{-1}+c_{-1})+
{\cal O}(1/r), -\frac{\i\csc\theta}{\sqrt{2}}(\cbar_{-1}-c_{-1})+
{\cal O}(1/r)
 \right),
\end{equation}
where 
\begin{eqnarray}
&&a_{-1}=\frac 12\sum^1_{m=-1}\left\{\overline{ A}_m+(-1)^m{A}_{-m}\right\} (_0Y_{1,m}), \\
&&c_{-1} =\sum^1_{m=-1}(-1)^{m+1}A_m\left( _{-1}Y_{1,-m}\right) , \\
&&\cbar_{-1} =\sum^1_{m=-1}\overline{A}_m\left( _1Y_{1,m}\right),
\end{eqnarray}
and the $A_m$ are arbitrary complex numbers. In the case of axial
symmetry, one has 
\begin{eqnarray}
a_{-1}=0, \\
c_{-1}=-\frac{\i\sin \theta}{\sqrt{2}}.
\end{eqnarray}

The other Killing vector compatible with a radiative gravitational
field is well known  to be the \emph{boost-rotation Killing vector} \cite{BicSch84,BicPra98}. Asymptotically, it reads:
\begin{equation}
\xi_2 ^\mu =(-u\cos \theta, r\cos \theta ,-\sin \theta ,0).
\end{equation}
Hence, in this case:
\begin{eqnarray}
c_{-1}=\sin\theta, \\
a_{-1}=\sqrt{2}\cos \theta.
\end{eqnarray}

Using these two Killing vectors, constraints on the 
diverse quantities of physical interest can be deduced \cite{BicSch84,BicPra98,Val00b}. The relevant results for us are the following:
\begin{eqnarray}
&&\sigma_2=-\csc(\theta)\left(K_1(w)+iK_2(w) \right), \\
&&\Re \,
\Psi_2^3=\frac{L(w)}{u^3}-\frac{1}{2u}\left(wK_1''(w)+2K_1'(w)\right), \\
&&\phi_2^1=\frac{1}{u^2}\left(H_1(w)+iH_2(w) \right), \\
&&\phi_1^2=\frac{1}{u^2}\left(G_1(w)+iG_2(w)\right)-
\frac{1}{\sqrt{2}u}\cot\theta\left(
H_1(w)+iH_2(w)\right),
\end{eqnarray}
where $w=\sin\theta/u$, and $K_1$, $K_2$, $L$, 
$H_1$, $H_2$, $G_1$, $G_2$ are arbitrary functions 
of the argument.  Regularity at the poles requires $G_1=G_2=0$ 
\cite{BicPra98}.

\section{The equations for $K_1$, $K_2$, $L$, $H_1$ and $H_2$}
As mentioned previously, we will only consider peeling
gravitational and electromagnetic fields, as polyhomogeneous
boost-rotation symmetric fields happen to be singular at the ``North''
and ``South'' poles \cite{Val00b}. Combining the evolution equations
for the different leading coefficients of the Maxwell and Weyl fields
together with the conditions of Petrov type D, the axial symmetry and
the boost-rotation symmetry one arrives to
 a system of 5
\emph{ordinary} differential equations for the functions $K_1$, $K_2$,
$L$, $H_1$ and $H_2$. These differential equations are highly coupled
and non-linear,  and far too lengthy to be displayed here.  In order
to get around the difficulty to extract some relevant physical
information out of them,  we adopt the approach of solving the
equations using formal expansions in $w=\sin\theta/u$. One might
expect these expansions to hold for $w$ close to $0$, i.e. for very
late times ($u>>1$).

\subsection{The simplest case: the C-metric}
How far one can go if one tries when trying an exact approach to the
equations for $K_1$, $K_2$, $L$, $H_1$, $H_2$? In order to answer
this, let us consider the easiest possible case in our analysis, that
of a type D, boost-rotation symmetric spacetime with hypersurface
orthogonal Killing vectors and no electromagnetic field. Under these
assumptions one obtains the following system of ordinary differential
equations for $K=K_1$ and $L$ which describe the news and mass aspect
of the C-metric:
\begin{equation}
3K''+4KK'+2wKK''+wK'''-2w^2L'-6wL=0
\end{equation}
and
\begin{eqnarray}
&&32(K')^3-8w^2(K')^2K'''+16w^2(K')^2L+16w^4KK'K''L-72w^4K'L^2-2w^4K''K''' +
\nonumber\\&&
24wK'^2K''+
4w^4K''L+10w^5K''L'-48w^5K'LL'+4w^5KK''L+
40w^3(K')^2L'
+\nonumber\\&&
40w^4K'K''L'+16w^3K'K''L-
8w^3K'K''K'''+16w^3K(K')^2L-2w^3(K'')^3-
\nonumber\\&&
24w^6K''LL'+12w^6K'''L=0.
\end{eqnarray}
These equations can be decoupled yielding 
a quadrature for $L$ and
the following fourth order 
differential equation for $K$:
\begin{eqnarray}
\bs&&10w^5K'K''K'''-10w^5KK''K^{(iv)}+14w^5K(K''')^2+60w^4K'(K'')^2+
48w^4KK''K'''+ 
\nonumber \\
\bs&&5w^4K''K^{(iv)}-7w^4(K''')^2-20w^4KK'K^{(iv)}+20w^4(K')^2K'''+144w^3K(K'')^2 
-19w^3K''K'''+\nonumber \\
\bs&&180w^3(K')^2K''+10w^3K'K^{(iv)}-72w^3KK'K'''-
42w^2(K'')^2+48w^2KK'K''+120w^2(K')^3+\nonumber \\
\bs&&46w^2K'K'''+66wK'K''+24wK(K')^2+48(K')^2=0, \label{fourthk}
\end{eqnarray}
which can be shown to have 3 three Lie point symmetries \cite{Ste89}, and therefore
can be reduced to the following Abel ordinary differential equation:
\begin{equation}
y'=\frac{2}{5}(224x^4+160x^3+22x^2-x)y^3+\frac{1}{5}(4x+7)y^2-
\frac{60x+8}{20x^2+5x}y.
\end{equation}
To the best of our knowledge this equation does not fall in any of the
known solvable categories of Abelian equations \cite{Sch99}. It is indeed
somehow frustrating having gone so far, and not being able to perform
the last step in the program! One may require a qualitative study of equation 
(\ref{fourthk}). This as well, is not an easy task due
to the highly nonlinear character of the fourth order differential
equation, which makes it very difficult to analyze the behavior of
the phase space close to critical points.

\subsection{The series solutions}
As discussed previously, we will attempt to extract some physical
information of the spacetimes using series expansions for $|w|<<1$, 
which describe the behavior of the system for late retarded
times. It is always possible to choose the size of this region
so that it will be completely contained in the radiative region 
of the spacetime  (the region above the ``roof''). The use of series 
carries some natural several drawbacks. In particular, it will not be
possible to  make any statement regarding the long
term behavior and existence of the spacetime. The exceedingly
complicated structure of the differential equations makes things
worse, not allowing us to make estimates of the convergence radius
or obtain the form of the general term in the expansion. Moreover,
 the coordinates $(u,\theta)$ are not well behaved at
$u=0$ (see the discussion in \cite{Val00b}).

\begin{figure}[tbt]
\centering \mbox{ \epsfig{file=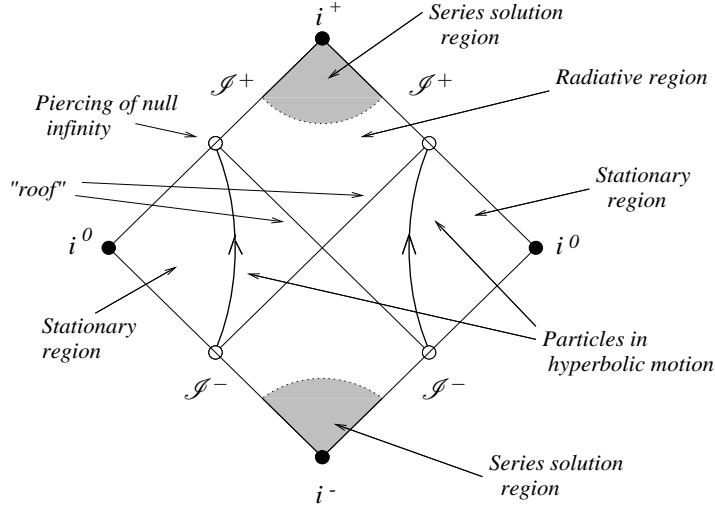,width=10cm,height=7cm}}
\put(-115,155){$\scri^+$}
\put(-110,40){$\scri^-$}
\put(-190,155){$\scri^+$}
\put(-200,40){$\scri^-$}
\caption{The different regions of a boost-rotation symmetric spacetime,
including the regions of validity of the series solutions.}
\end{figure}

If the regularity of the the solutions at the poles 
is demanded then the leading behavior of the series is given by:
\begin{eqnarray}
&&K_1=a_0+{\cal{O}}(w), \\
&&K_2=b_0+{\cal{O}}(w), \\
&&H_1=c_0+{\cal{O}}(w), \\
&&H_2=d_0+{\cal{O}}(w), \\
&&L=e_0+{\cal{O}}(w),
\end{eqnarray}
where $a_i$, $b_i$, $c_i$, $d_i$ and $e_i$ ($i\geq0$) are constants. 
The constant $a_0$ can be removed using a super-translation, so that 
in fact $K_1={\cal{O}}(w)$. The asymptotic shear transforms under super-translations according to \cite{Ste91}:
\begin{equation}
\tilde{\sigma}_{2,0}(\tilde{u},x^\alpha)=\sigma_{2,0}
(\tilde{u}-\alpha,x^\alpha)+\eth^2\alpha(\theta,\phi),
\label{supertranslatedsigma}
\end{equation}
and this being so  if
 initially one has $\sigma_2=-a_0\csc\theta+{\cal{O}}(w)$, then in order
 to get $\tilde{\sigma}_2={\cal O}(w)$ it is necessary to find a function
 $\alpha(\theta,\varphi)$ such that $\eth^2\alpha=a_0\csc\theta$.
 This is given by, 
\begin{equation}
\alpha(\theta,\varphi)=-2a_0\sin\theta,
\end{equation}
that despite its apparent simplicity defines a true super-translation\footnote
{This transformation is not just a translation  because the spherical 
harmonics expansion of the sine function consists of an infinite number
 of terms of the form $Y_{2n,0}$, $n=0,1,\ldots.$}. This choice
 of cuts of $\scri^+$ leads to the following series solutions:
\begin{eqnarray}
K_1(w)&&\bs=a_3\,{w^3} + a_5\,{w^5} + {{{\cal{O}}(w^7)}},\\
K_2(w)&&\bs=b_3\,{w^3} + b_5\,{w^5} + {{{\cal{O}}(w^7)}},\\
H_1(w)&&\bs=c_1\,w + \frac{3}{10\,\left( {{a_3}^2} + {{b_3}^2} \right)}
\left( 5\,(\,b_5\,
         \left( b_3\,c_1 + a_3\,d_1 \right)  + 
       a_5\,\left( a_3\,c_1 - b_3\,d_1 \right))\right.\nonumber\\
&&\bs  \;\;\;- \left.
        \left( a_3\,c_1 - b_3\,d_1 \right) \,
         \left( {{c_1}^2} + {{d_1}^2} \right)  \right) \,w^3
+{{{\cal{O}}(w^4)}},\\
H_2(w)&&\bs=w\,d_1 +\frac{3}{10\,\left( {{a_3}^2} + {{b_3}^2} \right)}
 \left( 5\,(a_5\,
         \left( b_3\,c_1 + a_3\,d_1 \right)  - 
        \,b_5\,\left( a_3\,c_1 - b_3\,d_1 \right))\right. \nonumber\\
&&\bs \;\;\;-  \left. \left( b_3\,c_1 + a_3\,d_1 \right) \,
         \left( {{c_1}^2} + {{d_1}^2} \right)
\right)\,w^3+{{{\cal{O}}(w^4)}},\\
L(w)&&\bs=4\,a_3 +\frac{2}{5}\, {{\left( 30\,a_5 - ({{c_1}^2} +
        {{d_1}^2}) \right) \,{w^2}}} + {{{\cal{O}}(w^4)}}.
\end{eqnarray}
The coefficients $a_3$, $b_3$, $a_5$, $b_5$ are free parameters of
 the gravitational field, whereas  $c_1$, $d_1$ play the same role 
for the electromagnetic field . In order to shorten the
 expressions of the components of the Weyl and Maxwell spinors, the
 following complex parameters will be introduced:
\begin{eqnarray}
&&q_1=c_1+id_1, \\
&&p_3=a_3+ib_3, \\
&&p_5=a_5+ib_5.
\end{eqnarray}
With this new notation, the leading terms of the components of
 the Weyl spinor are found to be
\begin{eqnarray}
\Psi_4^1=&&\bs6\,{\sin^2 \theta }
\left({{2\,{\bar p}_3}}\,u^{-5} + 
  {{5\,{\bar p}_5\,{{\sin^2 \theta }}}}\,u^{-7}\right)+
{\cal O}(u^{-9}),\\
\Psi_3^2=&&\bs-3\,{\sqrt{2}}\,\cos \theta\,\sin \theta \,
  \left( {{2\,{\bar p_3}}}\,u^{-4} +
    {{5\,\bar p_5\,{{\sin^2 \theta}}}}\,u^{-6} \right) +
{\cal O}(u^{-8}),\\
\Psi_2^3=&&\bs{{2\,\bar p_3\,\left( 2 - 3\,\,{{\sin^2 \theta }} \right) \,
      }u^{-3}} +
  {{{{\sin^2 \theta }}\,
      \left( 3\,\bar p_5\left( 4 - 5\,\,{{\sin^2 \theta}} \right) 
          - \frac{2\,q_1\,\bar q_1 }{5} \right) }}u^{-5}
+{\cal O}(u^{-7}),\\
\Psi_1^4=&&\bs 3\,{\sqrt{2}}\,\cos \theta \,\sin \theta \,
  \left(\bar p_3\,u^{-2} + 
    {{\frac{1}{20}\,\left( 5\,\bar p_5
           \left(1 - 5\,\cos2\theta  \right) \,
            +4\,q_1\,{\bar q_1} \right)\,u^{-4} }} \right)+{\cal O}(u^{-6}),\\
\Psi_0^5=&&\bs3\sin^2\theta\left(\,\pbar_3\,u^{-1} -\frac{1}{20}
        \left( 5\pbar_5\left( 3 + 5\,\cos2\theta \right) \,
          - 8\,q_1\,\bar q_1 \right)u^{-3}\right) +\O(u^{-5}), \\
\Psi_0^6=&&\bs-\sin^2\theta\left(\frac{15}{2}{{\pbar_3}} -\frac{3}{8}
  {{\,
      \left( 5\pbar_5\left( 5 + 7\,\cos 2\theta \right) \,
          - 8\,q_1\,\bar q_1 \right)u^{-2} }}\right) +\O(u^{-4});
\end{eqnarray}
whereas the relevant terms of the components of the Maxwell spinor are
\begin{eqnarray}
\phi_2^1=&&\bs q_1 \sin\theta\left({{{u^{-3}}}} +
  {\frac{3}{10\,\bar p_3}{{{\sin^2 \theta}}\,
      \left( 5\,{p_5} - q_1\,{\bar q_1} \right) }}u^{-5}\right)+
{\cal O}(u^{-7}),
\\
\phi_1^2=&&\bs-{\frac{q_1\cos \theta }{{\sqrt{2}}}}\left(u^{-2} -
 \frac{3}{10\,\bar p_3} {{{{\sin^2 \theta }}\,
      \left( 5\,{\bar p_5} - q_1\,{\bar q_1} \right) }}u^{-4}\right)+
{\cal O}(u^{-6}),\\
\phi_0^3=&&\bs-\frac{q_1\,\sin \theta }{2}
\left( u^{-1} - \frac{1}{10\,\bar p_3}
  {{\left( 2 - 3\,{{\sin^2 \theta}} \right) \,
      \left( 5\,{\bar p_5} - q_1\bar q_1 \right) }}
u^{-3}\right)+{\cal O}(u^{-5}), \\
\phi_0^4=&&\bs\frac{3\,q_1\,\sin\theta}{4}\left(
1 -   {\frac{1}{10\,\bar p_3}
      \left(4 - 5\,{{\sin^2 \theta}} \right) \,
      \left( 5\,{\bar p_5} - q_1\bar q_1
 \right)u^{-2} }\right)+{\cal O}(u^{-4}).
\end{eqnarray}
It is worth noting the existence of a peeling off property of the
 $u$ dependence of these coefficients: 
\begin{eqnarray}
&&\Psi_n^{5-n}=\O(u^{-1-n}), \\
&&\phi_m^{3-m}=\O(u^{-1-m}),
\end{eqnarray}
with $n=0,\ldots,4$ and $m=0,\ldots,2$. This behavior can be
understood in the following way: one can perform a construction
equivalent to the one we presented, but for past-oriented light cones,
parametrized by an advanced time $v$ coordinate. The components of the
Weyl and Maxwell spinors will peel off with  respect to the affine
parameters of the null generators of these past-oriented cones (recall
that boost-rotation symmetric spacetimes are time symmetric!). Now,
close to $\scri^+$, one can show that $r_v=u+\O(u^{-1})$, where $r_v$
is the affine parameter in the past-oriented system. Similarly, $v\sim
r$ in a neighbourhood of $\scri$, and $l^a$ and $n^a$ are ``almost''
parallel to $n^a_v$ and $l^a_v$ respectively. Thus, the leading
coefficients should peel off with respect to $r_v$. Whence, they also
have to peel off with $u$.

\section{Asymptotic and radiative properties}
We have already gathered enough information
so as to  study some the spacetime's physical 
properties at $\scri^+$
in the late-time regime. 
\subsection{The electromagnetic charge}
The total charge in the spacetime is given by
\begin{eqnarray}
e=\int_{S^2}\phi_1^2(_0Y_{0,0}) \d S 
 =\sqrt{\pi}\int_{S^2}\phi_1^2 \sin \theta \d \theta 
 = O(u^{-5}),
\end{eqnarray}
i.e. it vanishes up to $O(u^{-5})$. Moreover, since the total electromagnetic
 charge in the spacetime is a conserved quantity, it should not
 contain any $u$ dependence. Therefore one can affirm that for our
 spacetimes 
\begin{equation}
e=0.
\end{equation}
This is in agreement with the interpretation 
put forward by Cornish \& Utteley
 \cite{CorUtt95a,CorUtt95b}, who interpreted the charged C-metric as
 two charged black holes of opposite charges in hyperbolic motion.

\subsection{The Bondi mass and the NUT mass}

Of great physical relevance is the study of the mass loss of the system
 due to radiative processes. This can be done by evaluating the
 \emph{Bondi mass}, which is given by:
\begin{equation}
M_B=-\frac{1}{2}\int_{S^2}\mbox{Re\,}\left( \Psi_2^3 +\sigma_2
 \dot{\sigmab}_2 \right ) (_0,Y_{0,0})\d S,
\end{equation}
and can be shown to be non-increasing. Another related quantity, the 
NUT mass \cite{RamSen81,AshSen82} will also be  of some interest due
 to the generic existence of nodal singularities in the interior of
 boost-rotation symmetric spacetimes \cite{BicSch89a}. The NUT mass
 is essentially an imaginary version of the Bondi mass:
\begin{equation}
M_{NUT}=-\frac{1}{2}\int_{S^2}\mbox{Im\,}\left( \Psi_2^3 +\sigma_2 
\dot{\sigmab}_2 \right ) (_0,Y_{0,0})\d S,
\end{equation}
The evaluation of these integrals for our solutions yields:
\begin{eqnarray}
\bs\bs&& M_B=4\,\sqrt{\pi}\,\left(\frac{1}{15}\,{q_1{\bar q_1}}{{u^{-5}}}+
\frac{2}{5}\,{{p_3\,{\bar p}_3}{{u^{-7}}}} 
 - 
  \frac{16}{35}\left(p_5\,{\bar p_3} + 
        p_3\,{\bar p_5} \right)\,{{u^{-9}}}
+{\frac{32}{63}\,{p_5\,{\bar p}_5}\,
{{u^{-11}}}}
 \right)+{\cal O}(u^{-13}) \label{bondimass}\\
\bs\bs&& M_{NUT}=\frac{16\sqrt{\pi}}{35}\left(p_3\pbar_5-\pbar_3p_5\right)u^{-9}+\O(u^{-13}). \label{nutmass}
\end{eqnarray}
If the electromagnetic field is absent  then,
\begin{equation}
M_B=\frac{8\sqrt{\pi}}{5}p_3\pbar_3\,u^{-7}+\O(u^{-9}),
 \label{nochargebondimass}
\end{equation}
thus $m=p_3\pbar_3=|p_3|^2$ can be interpreted as the mass of 
the system at a fiduciary retarded time (say $u=1$). Note as well
 that if $u\!\longrightarrow\!\infty$ then $M_B\!\longrightarrow\!0$, i.e.
 all the mass in the spacetime is carried away. This is consistent
 with the standard interpretation of boost-rotation symmetric 
as spacetimes describing particles in hyperbolic motion. These particles
 eventually pierce null infinity, thus leaving the (unphysical)
spacetime.
 Consistent with these results, Dray has shown \cite{Dra82} that the
 ADM mass of the C-metric is zero. His analysis shows that the reason
 for this is the particular kind of isometries this spacetime has. 
Therefore, it is very likely that this result also holds for the whole
 class of boost-rotation-symmetric spacetimes.

A similar discussion can be done with the NUT mass. The constant
 $g=p_3\pbar_5-\pbar_3p_5$ can be interpreted as the measure of 
the ``strength'' of the nodal singularities joining the particles
 in hyperbolic motion at a fiduciary time, $u=1$.

\subsection{The Newman-Penrose constants}
Further insight on the physical meaning of the parameters $q_1$ and
$p_3$ can be achieved by looking at the Newman-Penrose constants of
the gravitational and electromagnetic field.

As shown by Exton et al. \cite{ExtNewPen69}, there will be two sets of 
conserved quantities for the Einstein-Maxwell system under
consideration: 
one for the electromagnetic field only, and one for the combined 
electromagnetic-gravitational system. The electromagnetic NP constants 
are given by:

\begin{equation}
\mathcal{F}_m=\int_{S^2}\phi_0^4 (_1\Ybar_{1,m})\, \d S\,.
\label{electromagneticnp}\end{equation}
Note that due to the axial symmetry, all the constants, except
for the one corresponding to  
$m=0$, will vanish identically  (the $m=0$ spherical harmonic has
 no $\varphi$ dependence).

The NP constants for the combined gravitational-electromagnetic system
 are somewhat more complicated as they involve the 
inverse operator $\eth^{-1}$:

\begin{equation}
\mathcal{G}_m=\int_{S^2}\left\{ \Psi_0^6+
4\phi_0^1\ethbar^{-1}(\phib_1^2-\overline{E})-
4\overline{E}\ethbar^{-1}(\phi_0^4-F) \right \}(_2\Ybar_{2,m})\, \d S\,,
\label{gravitoelectromagneticnp}\end{equation}
where
\begin{eqnarray}
&&E= e (_0Y_{0,0}), \\
&&F= \sum_{m=-1}^{1} \mathcal{F}_m(_1Y_{1,m}),
\end{eqnarray}
 and
\begin{eqnarray}
&&\ethbar^{-1}(\phib_1^2-\overline{E})=
-\sum_{l=1}^\infty \sum_{m=-l}^l\frac{(_1Y_{l,m})}{\sqrt{l(l+1)}}
 \int \phib_1^2 (_0\Ybar_{l,m})\, \d S\,, \\
&&\ethbar^{-1}(\phib_0^4-\overline{F})=
-\sum_{l=2}^\infty \sum_{m=-l}^l\frac{(_2Y_{l,m})}{\sqrt{(l-1)(l+2)}}
 \int \phib_0^4 (_1\Ybar_{l,m})\, \d S\,.
\end{eqnarray}

The spacetimes under
discussion are axially symmetric, and thus they only have one nonzero
complex electromagnetic and one complex gravitational NP constants
(those corresponding to the $m=0$ spherical harmonics). In the region below the ``roof'', the
spacetime is stationary, and therefore its electromagnetic
Newman-Penrose constants will be of the form (mass
monopole)$\times$(electric dipole)$-$(electric charge)$\times$(mass
dipole), while the gravitational constants will be of the form (mass
dipole)$^2-$(mass monopole)$\times$(mass quadrupole) \cite{ExtNewPen69}. 

Using equation (\ref{electromagneticnp}) one finds that the
electromagnetic 
NP constant is given by: 
\begin{equation}
\mathcal{F}_0=\sqrt{\frac{3}{8\pi}}\,q_1.
\end{equation}
Therefore $q_1$ is a product of a mass monopole and electric dipole,
 for there is no electromagnetic charge in the spacetime. Similarly,
 using equation (\ref{gravitoelectromagneticnp}) the evaluation of the
 gravitational NP constants yield:
\begin{equation}
\mathcal{G}_0=-6\sqrt{\frac{5}{\pi}}\,\pbar_3.
\end{equation}
Thus, $p_3$ has to be interpreted as a quantity of quadrupolar nature,
 even if strictly speaking the spacetime does not has a mass monopolar
 moment for the particle undergoing hyperbolic motion in a
 boost-rotation symmetric spacetime does not always are present in
 pairs.

Furthermore, let us recall the formulae for radiated power for
 the electromagnetic and the linearized gravitational field,
 the so called \emph{dipole and Einstein's quadrupole formulae}
 \cite{LanLif51}:
\begin{eqnarray}
&&I_{elect}\propto\, \ddot{\bf\mbox{\bf d} }^2, \\
&&I_{grav} \propto\, \stackrel{\ldots}{\mbox{\bf D}}^2,
\end{eqnarray}
where {\bf d} is the dipolar moment of the charge distribution giving
 rise to the radiation field, and {\bf D} is the quadrupolar
 moment of the mass distribution (a 3-dimensional tensor). 
>From the Bondi mass one can obtain by differentiation 
the flux of energy through null infinity. This is,
\begin{equation}
\dot{M}_B=-4\,\sqrt{\pi}\,\left(\frac{1}{3}\,{q_1{\bar q_1}}{{u^{-6}}}+
\frac{14}{5}\,{{p_3\,{\bar p}_3}{{u^{-8}}}}\right)+\O(u^{-10}).
\end{equation}
Therefore the spacetimes under discussion radiate according to the
 dipole and Einstein's quadrupole formula up to the leading terms.
 Higher order corrections are due to the damping of the
 electromagnetic
 field by the gravitational field, some other nonlinear effects like
 gravitational wave backscattering. Note as well how the flux
 of energy due to the gravitational field decays much faster than the
 flux due to the electromagnetic field.
 Thus at late times the main contribution to mass loss is
 electromagnetic
 in origin (Poynting vector).

\subsection{Polarization states of the electromagnetic and
 gravitational waves}

The discussion along this paper has focused on peeling electromagnetic
and gravitational fields, and consequently the fields at
null infinity are well defined.  If one wishes to study the
states of  polarization of the electromagnetic or the gravitational
fields, it is only necessary to consider the behavior of the  type N part
of the fields, that is, the leading coefficients of $\phi_2$ and
$\Psi_4$ (i.e. $\phi_2^1$ and $\Psi_4^1$). The spinorial electromagnetic field is related to the components of
the Maxwell field tensor via:
\begin{eqnarray}
&&E_1-iB_1=\phi_0-\phi_2 \label{EB1} \\ 
&&E_2-iB_2=i(\phi_0+\phi_2) \label{EB2}   \\
&&E_3-iB_3=-2\phi_1 \label{EB3}
\end{eqnarray}
In the case here the  electromagnetic field can  be seen to have two
polarization states,  associated with the real part and imaginary
parts  of $\phi_2^1$ respectively. One way to see this is to realize
that if $d_1$ is set to zero then  there will still be  non vanishing
components of the electromagnetic field. A similar phenomenon happens
if $c_1$ is set to zero. This is nothing but  a consequence of the
existence of two different polarization  states. If $\phi_2^1$ is real,
then the electric and magnetic fields ---see equations
 (\ref{EB1})-(\ref{EB3})--- lie along the $e_1$  and
$e_2$  directions respectively. Note that
$e_1$ and $e_2$ are orthonormal
 vectors transverse to the null generators of 
outgoing light cones, explicitly:
\begin{eqnarray}
&&e_{2}^a=\frac{1}{\sqrt{2}}i(m^a-\mbar^a), \label{ortho3}\\
&&e_{3}^a=\frac{1}{\sqrt{2}}(l^a-n^a) \label{ortho4}.
\end{eqnarray}
Both 3-vectors
are orthogonal, and have the same norm, as  it should be expected from
a plane wave. Let us call  this polarization state $P_x$ (the
polarization vector lies on the $e_1$ direction ---x axis---
\cite{Ste82}). When $\phi_2^1$ is purely imaginary ($P_y$ polarization), the
situation gets reversed and the electric field and magnetic 
lie along the $e_2$  and $e_1$ directions respectively (see figure 6.3).
\begin{figure}[tbt]
\centering \mbox{ \epsfig{file=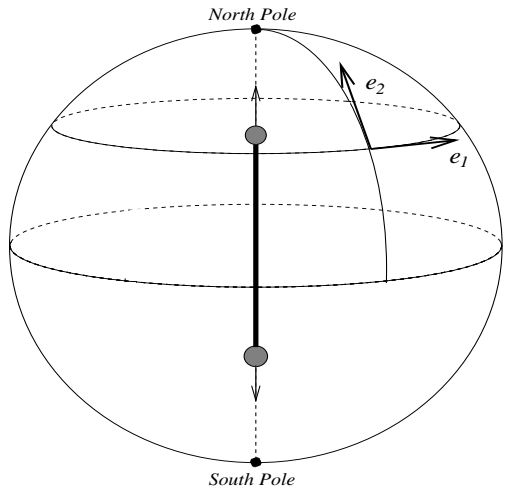,width=7cm,height=7cm}}
\caption{The directions of the polarization states of the
electromagnetic
 and gravitational fields with respect to the celestial sphere.}
\end{figure}
By calculating the Poynting vectors for each state it can be  checked
that each  configuration carries energy independently from the other,
supporting again the assertion made on the existence of two
independent degrees of freedom for the electromagnetic field.

For the gravitational field, the situation is fairly similar to that of 
 the electromagnetic field. Again, there are two set of
parameters: $\{a_3,a_5\}$ (the real parts of $p_3$ and $p_5$),
 and $\{b_3,b_5\}$ (the imaginary parts of $p_3$ and $p_5$).
 Both of them play an equivalent role in the expression for the
 Bondi mass. Hence both states of the gravitational field carry
 energy independently, i.e. they are true degrees of freedom
 (polarization states) of the gravitational field. The one 
associated to the real part of the Weyl tensor, which is given by
 the $a$-parameters, describes a $(+)$ polarization state. This
 can be seen by considering the geodesic
deviation equation \cite{Sze65}:
\begin{equation}
\delta \ddot{x}^a=\frac{1}{2} \mbox{Re\,}\Psi_4^1 \left (e_2^a
e_{2c}-e_1^ae_{1c} \right ) \delta x^c, \label{geodesic}
\end{equation}
therefore there will be tidal forces along the
$e_1$ and $e_2$ directions. Similarly, if one considers the
polarization
 state associated to the imaginary part of the Weyl tensor, one can
 perform a spin boost so that $\Psi_4$ becomes real, so that one can
 use again equation (\ref{geodesic}). This polarization state
 will be a $(\times)$ one, with tidal forces rotated $\pi/4$ with
 respect to the $e_1$ and $e_2$ directions.

Note that if the Killing vectors are hypersurface orthogonal, then
 there is only one polarization state of the gravitational field
 ($+$).
 By contrast, the electromagnetic field can always have two
 polarizations
 states, as long as $q_1$ is complex (a magnetic
 dipole moment is then present!).

\subsection{Conclusions}
The main objective of this paper was to construct 
 spacetimes  that could be used as examples and test bench of different techniques and ideas used in the description of
 on asymptotic and radiative properties of the gravitational and electromagnetic fields.  

Boost-rotation symmetric spacetimes are regarded as the best
suited candidates for such a test bench since they contain the
 only examples known up to date  of asymptotically flat
 radiative exact solutions. Perhaps the most 
remarkable result  was that of showing 
how the Newman-Penrose constants allow one to consider, at least  up to
quadrupoles,
multipolar moments of non-stationary
 spacetimes without having to resort to the linearized 
theory (cfr. the results on the mass loss). As seen, the complexity of the equations involved precludes the obtention of nice closed expressions; however, the asymptotic expressions for late times have allowed to extract most of physical properties of interest; this regardless of the natural limitations that this approach carries.

\end{document}